\documentclass[10pt,conference,compsocconf,letterpaper]{IEEEtran}

\usepackage{amsmath}
\usepackage{amssymb}
\usepackage[vlined]{algorithm2e}
\usepackage{color}
\usepackage{graphicx}
\usepackage{url}
\usepackage{subfig}

\newtheorem{definition}{Definition}[section]

\newcommand{\prism}{PriSM}

\usepackage{xcolor}

\newcommand{\remove}[1]{}

\title{PriSM: A Private Social Mesh for Leveraging Social Networking at Workplace}

\author{\IEEEauthorblockN{Stefano Braghin}
\IEEEauthorblockA{Digital Enterprise Research Institute\\
National University of Ireland, Galway\\
Ireland\\
\texttt{stefano.braghin@deri.org}}
\and
\IEEEauthorblockN{Jackson Tan, Rajesh Sharma and Anwitaman Datta}
\IEEEauthorblockA{School of Computer Engineering\\
Nanyang Technological University\\
Singapore\\
\texttt{\{jacktty,raje0014,anwitaman\}@ntu.edu.sg}
}
}

\begin{document}

\maketitle

\begin{abstract}
  In this work we describe the \prism{} framework for decentralized
  deployment of a federation of \emph{autonomous social network}s
  (ASN). The individual ASNs are centrally managed by organizations
  according to their institutional needs, while cross-ASN interactions
  are facilitated subject to security and confidentiality requirements
  specified by administrators and users of the ASNs. Such
  decentralized deployment, possibly either on private or public
  clouds, provides control and ownership of information/flow to
  individual organizations. Lack of such complete control (if third
  party online social networking services like Facebook or Yammer were to be used) has so far
  been a great barrier in taking full advantage of the novel
  communication mechanisms at workplace that have however become
  commonplace for personal usage with the advent of Web 2.0 platforms
  and online social networks. \prism{} provides a practical solution
  for organizations to harness the advantages of online social
  networking both in intra/inter-organizational settings, without
  sacrificing individual as well as organizational autonomy, security and confidentiality needs.

\end{abstract}

\begin{IEEEkeywords} online social networking platform; decentralization; federation; autonomy; private/public cloud; workplace
\end{IEEEkeywords}

\maketitle

\section{Introduction}
\label{sec:intro}
Online social networking and other Web 2.0 applications have brought
in a paradigm shift in the manner in which people communicate and
interact online. Realizing the versatility, flexibility and reach of
open online social networks such as Facebook and Twitter, they have
been widely embraced by organizations for public relations as well as
marketing and monitoring purposes. Some relevant Web 2.0 technologies,
such as Wikis are also readily deployed within corporate
Intranets. However, other platforms, particularly social networking,
despite its preponderance in the Internet setting, are yet to become
an integral part of individual organizations' internal communication
and workflow infrastructure.

While the new modes and (more importantly) opportunities of
interaction that social networking platforms provide can
significantly help improve an organization's internal dynamics,
there have so far been several barriers in wide-scale adaption of
such infrastructure in workplace. Foremost, a Facebook like platform
which is open to all, or even a more closed system like Yammer, hosted and controlled by a third party, is
unsuitable for storing and communicating sensitive business data and
information. In contrast to Wiki-engines which can be privately
deployed, there has been a relative lack of out-of-the-box social
networking platform software.\footnote{We note that recent
implementations arising from works on decentralized online social
networks are partially filling up the void.} Furthermore, each
organization is differently structured, and carries out distinct
activities, thus it is essential to be able to map these
organizational structures and processes in the platform. Ultimately,
even if most of the interactions are carried out within corporate
boundaries, ability to interact with outside entities, for example,
with customers or suppliers (or even across different departments or
project groups within same organization), requires mechanisms
enabling easy and flexible ways to express rules of engagements and
enforce and monitor the same subject to various security and
confidentiality needs of all the stake-holders --- particularly that of the organizations and the individuals.

In this paper we present a framework for deploying autonomous social
networks (ASNs) that can be run and administered independently, and
can further be federated with other ASNs through trusted peering
links. \prism {} (Private Social Mesh) implements such a framework.
This results in a hybrid architecture, where individual ASNs follow traditional OSN's client-server model, while the federation is achieved in a peer-to-peer manner.

An analogy for such decentralized social networking platform
deployment may readily be drawn from the way `emails' work.
Individual organizations often choose to run their own private email
servers catering to their users, while these users can also
communicate with users using other email services. Furthermore,
organizations may also choose to rent the server functionalities or
even the whole email service from a cloud based service provider.
Our \prism{} implementation allows similar deployment models, i.e. deployed from scratch on personal servers/private clouds or on a public cloud service providing Infrastructure as a Service (IaaS), or alternatively, get administrative access to a preinstalled configurable instance, akin to Software as a Service (SaaS).
However, in contrast to email's any-to-any communication paradigm,
\prism{} allows ASN administrators as well as an user's superiors
from within the organizational hierarchy to determine intra/inter ASN communication restrictions. Thus, from an operational point of view, \prism{} provides bottom up access control where individuals determine which other users have access to specific resources owned by that user, as in traditional online social networks, but it also allows top-down access control, where (sub-)domain administrators/delegates determine the rights and rules determining the possible actions that individuals can carry out. From an infrastructural perspective, \prism{} makes similar trust assumptions as typical email server deployments, and each organization (administrators) orchestrates the data storage and flow within individual ASNs. 

The main contributions of this work are as follows: (i) A framework
for decentralized deployment of autonomous social networks (ASNs) which allow
their users to map their respective organizational structures and
processes such as departments, project (sub-)groups, etc. is
proposed. (ii) The proposed framework supports federation of ASNs with peering mechanisms
for inter-ASN user interactions. (iii) It allows to scope
intra/inter-ASN interactions flexibly, determined by users (and
their superiors) subject to business as well as individual privacy
and confidentiality needs. Furthermore, the decentralized architecture naturally allows for deployment of individual ASNs in both private/public server/cloud environments.

The rest of the paper describes the model and implementation of the \prism{}
framework as follows: Section~\ref{sec:model} describes the
network model while Section~\ref{sec:ac} describe the ``frontier'' information
propagation \& control mechanism. In Section~\ref{sec:privileges} we present the access control mechanism deployed in \prism{} and Section~\ref{sec:architecture} presents
the architecture of the framework detailing encountered implementation issues and summarizing lessons learnt from the experience. Relevant related works are discussed in Section~\ref{sec:related}. We draw our conclusions and outline our ongoing and planned extensions of \prism{} in Section~\ref{sec:conclusion}.


\section{The social mesh model}
\label{sec:model}

We define a social mesh as a network of social networks, described
next by borrowing some terminologies from sociology
literature~\cite{AMET:AMET219}.  Naturally, the model is rather
standard besides the different user groups required by PriSM's model.
Note that in what follows, we assume that a user is employed only in a
single organization. Hence, in our model, an individual with multiple
accounts across different ASNs is considered as distinct users.

\begin{figure}[htb]
\centering
\includegraphics[width=\columnwidth]{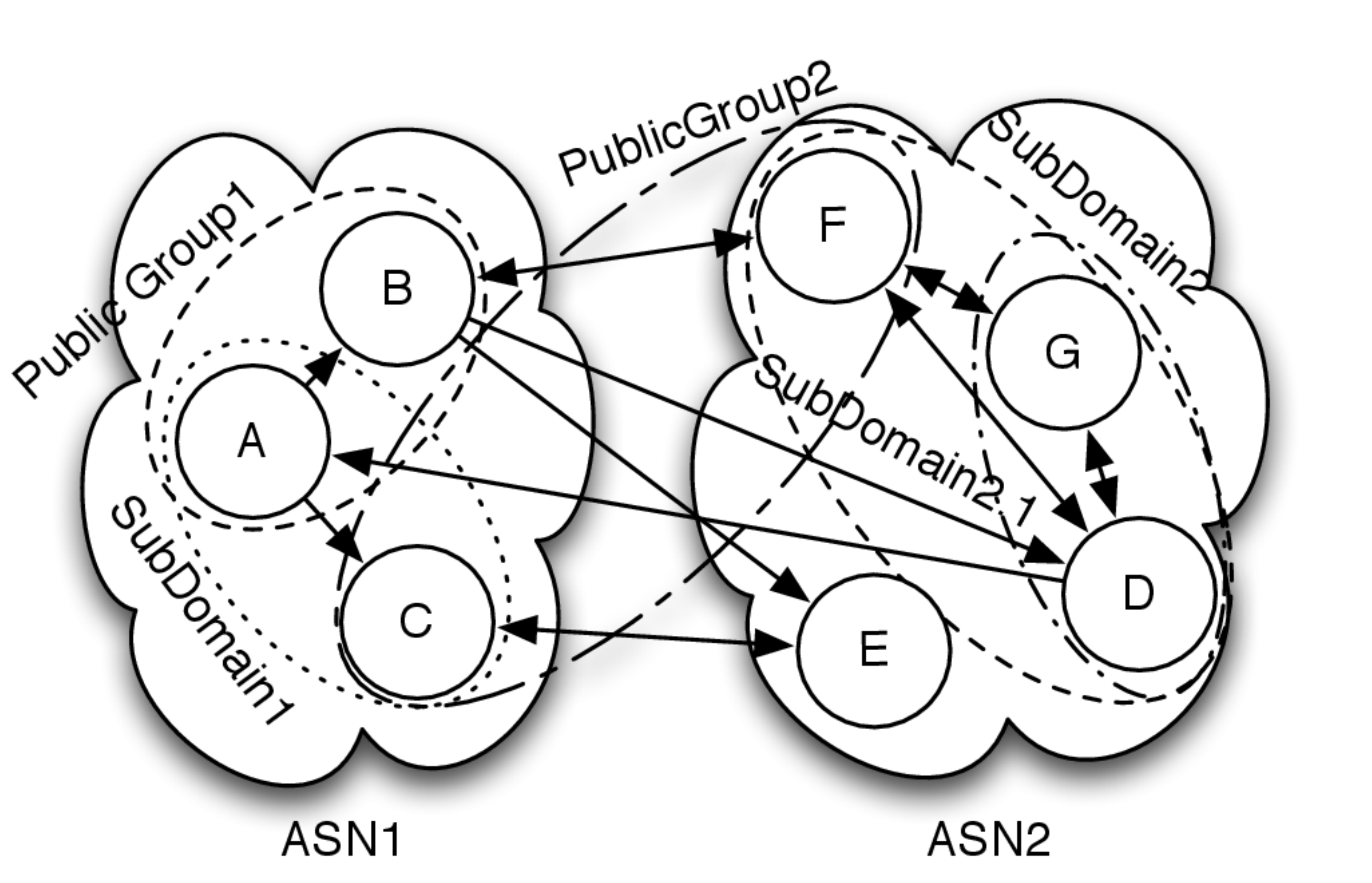}
\caption{The social mesh model.}
\label{fig:model}
\end{figure}

Figure~\ref{fig:model} shows a simple example instance of a Social Mesh.

As briefly mentioned in Section~\ref{sec:intro}, \prism{} models what
we call a \emph{Social Mesh}, which is a network interconnecting
distinct \emph{Autonomous Social Networks} (or ASN for short).

\begin{definition}[Social Mesh]
\label{def:mesh}
A \emph{Social Mesh} $SM$ is a tuple $$\left< \mathcal{ASN},
  \mathcal{U}, \mathcal{PG} \right>$$ where $\mathcal{ASN}$ is the set
of autonomous social networks and $\mathcal{U}$ is the set of users.
Each user $u \in \mathcal{U}$ belongs to exactly one autonomous social
netowk $asn \in \mathcal{ASN}$.  Finally, $\mathcal{PG}$ is the set of
public groups defined in $SM$.
\end{definition}

An ASN is the social network defined within a given
organization. Thus, it defines the members, their privileges and the
communication channels which are existing within the organization. The
ASN also enforces the organization's policies in terms of information
flow, beside the users' ones.

\begin{definition}[Autonomous Social Network]
\label{def:administrative_domain}
Given a social mesh $SM$, an autonomous social network $asn \in
\mathcal{ASN}(SM)$ is a tuple of the form $$\left<a, \mathcal{UD},
  \mathcal{SD}, rsd, 
  \mathcal{R}
\right>$$ where $\mathcal{UD} \subset \mathcal{U}(SM)$ is the set of
users of $asn$ and $a \in \mathcal{UD}$ is the administrator of the
autonomous social network.  Moreover, $\mathcal{SD}$ is the set of
subdomains defined in $asn$, $rsd \in \mathcal{SD}$ is the main
subdomain and 
$\mathcal{R}$ is the set of roles defined in
$asn$.
\end{definition}

The information flow across different users of an ASN and across
different ASNs is managed by means of \emph{circles}.  A circle is a
group of users of the social mesh and an associated set of rules
controlling how information -- messages -- associated to such circle
can be accessed by users not belonging to the circle itself.  Thus, we
assume that, in general, information associated to a circle are
accessible by members of the circle.  In \prism{}, one may associate
different circles to a message by means of so called \emph{tag set}
(denoted by $\mathcal{T}(m)$\footnote{We use the notation $P(O)$ to
  refer to the property $P$ of the object $O$}), which is the set of
circles controlling who is entitled to access the message, and a
\emph{conflict set} (denoted by $\mathcal{I}(m)$), which is the set of
circles whose members cannot access the message.

\prism{} allows the specification of different types of circles to
represent the different types of users' groups existing within real
world organizations.  Because of that, we need circles representing
both the internal structure of complex organizations, as well as other
circles not directly mapping formal structure of an organization.  We
call circles materializing structures of an organization as
\emph{subdomains}.

\begin{definition}[Subdomain]
  \label{def:subdomain}
  Given a social mesh $SM$ and an autonomous social network $asn \in
  \mathcal{ASN}(SM)$, a subdomain $sd \in \mathcal{SD}(asn)$ is a
  tuple of the form $$\left< n, \mathcal{M}, \mathcal{PR}, \mathcal{P},
    a, f\right>$$ where $n$ is the identifier of the subdomain,
  $\mathcal{M} \subseteq {UD}(asn)$ is the set of members of $sd$,
  $\mathcal{PR}$ is the set of privileges granted by the administrator
  to the members of the subdomain $sd$, $\mathcal{P}$ is the set of
  rules defining the constraints a user must satisfy in order to
  access messages tagged with $sd$ (or any circle which is a child of
  $sd$, see Definition~\ref{def:subdomain_hierarchy},
  Definition~\ref{def:public group_hierarchy} and
  Definition~\ref{def:collection_hierarchy}) and $a \in
  \mathcal{M}(sd)$ is the administrator of $sd$.
\end{definition}
More than one user may have administrative privileges on a given
subdomain $sd$, as we discuss in Section~\ref{sec:privileges}. $a(sd)$
is the user who initially received the charge of managing $sd$. We
assume that $a(sd)$ does not change over time even if administrative
privileges may be granted and revoked to other users. Thus, we assume
that $a(sd)$ will always be granted administrative privileges over
$sd$.

Example of subdomains may be departments of a university or branches
of a company.  Subdomains are organized in a hierarchy representing
the parent-child relationship existing among the different departments
of the
company.
\prism{} does not restrict the number of children of a subdomain, on
the other hand, we restrict the number of father of a subdomain to at
most one. Similarly, we will later assume similar restrictions to
circles. The reason is related to how \prism{}'s information
propagation mechanism works, see Section~\ref{sec:ac}.

\begin{definition}[Subdomain hierarchy]
\label{def:subdomain_hierarchy}

Let $\mathcal{SD}$ be the set of subdomains defined for an autonomous
social network $asn$. The subdomain hierarchy $$\phi_\mathcal{SD} :
\mathcal{SD} \rightarrow \mathcal{SD} \cup \{\bot\}$$ is the function
defining the hierarchy among the subdomains.

\end{definition}

The root of the subdomains' hierarchy is called \emph{main
  subdomain}. Such subdomain represents the organization
itself. Therefore, it is required to be defined and unique.

\begin{definition}[Main subdomain]
  Given a set of subdomains $\mathcal{SD}$ for a given autonomous
  social network $asn$ we define as \emph{main subdomain} for $asn$
  the subdomain $msd \in \mathcal{SD}$ such that
  $\phi_\mathcal{SD}(msd) = \bot$.  We further assume that $msd$
  exists and is unique for each $asn$, which means that $\forall sd
  \in \mathcal{SD}$ if $\phi_\mathcal{SD}(sd) = \bot$ then $sd = msd$.
\end{definition}

On the other hand, circles representing groups created for official
purposes, but without a direct mapping into the organization's
structure, are called \emph{public groups}.

\begin{definition}[Public Group]
  \label{def:public group}
  Given a social mesh $SM$, a public group $PG \in \mathcal{PG}(SM)$
  is a tuple of the form $$\left<o, \mathcal{M}, \mathcal{B},
    \mathcal{P}
  \right>$$ where $o \in \mathcal{U}$ is the user who created the
  public group, $\mathcal{M} \subseteq \mathcal{U}$ is the set of
  users who are member of $c$.  $\mathcal{B} \subseteq \mathcal{M}$ is
  the set of ``bosses'' of $c$, i.e. the users who can modify
  $\mathcal{P}$, the set of rules associated to the public group.
\end{definition}

As an example, a public group may be a team of physicians and nurses
working on a specific disease. The different cases related to that
disease may be handled by users belonging to different departments of
the hospital, such as users from the Cardiology Department (a
subdomain) and users from the Elderly Service Department (another
subdomain). Hence, the main feature characterizing a public group is
the purpose for which it has been created. Some ASNs may allow users
to create and join public groups created for purposes not directly
work-related, such as a group created to simplify the communication
among the players of the Nurse's Soccer Team. As opposed to
subdomains, members of a public group may belong to different ASNs,
such as a research project carried out by researchers and professors
from different universities.  Similarly to subdomains, public groups
may be organized hierarchically. More precisely, in \prism{}, a public
group may specify a public group or a subdomain as parent.

\begin{definition}[Public Group hierarchy]
\label{def:public group_hierarchy}
Let $\mathcal{PG}$ be the set of public groups and let $\mathcal{SD}$
be the set of subdomains of a social mesh $SM$. The public group
hierarchy $$\phi_\mathcal{PG} : \mathcal{PG} \rightarrow \mathcal{PG}
\cup \mathcal{SD} \cup \{\bot \}$$ is the function defining the
parent-child relationship for public groups.
\end{definition}

Finally, \prism{} allows users to define personalized circles called
\emph{private groups} in which users are categorized according to the
preferences of the creator of the circle.
\begin{definition}[Private Group]
\label{def:collection}
Given a user $u$, a private circle $prg \in \mathcal{PRG}(u)$ is a
couple of the form $$\left<\mathcal{M}, \mathcal{P} 
\right>$$ where $\mathcal{M} \subseteq \mathcal{U}$ is the set of users
who are member of $prg$.  $\mathcal{P}$ is the set of rules associated
to the private group.
\end{definition}

Again, private groups can be organized hierarchycally but only among
private groups of the same creator.

\begin{definition}[Private Group hierarchy]
\label{def:collection_hierarchy}

Let $\mathcal{PRG}(u)$ be the set of private gropus for a given user $u$.
The private group hierarchy
$$\phi_\mathcal{PRG}^u: \mathcal{PRG}(u) \rightarrow \mathcal{PRG}(u) \cup \{\bot\}$$
is a function defining the parent-child relationship among the private groups of $u$.
\end{definition}

Such private groups are strictly private to the creator of the circle, and thus unknown to the users who are categorized.
Private groups provide a tool to control the flow of an individual's messages in a fine-grained manner (akin to the use of circles in Google+), for example specifying that a message is visible only to the user categorized to a specific private circle.

As a concrete example, consider a physician working on a very
sensitive case.  She/he may create a private group of ``untrusted
colleagues'' to avoid such users from receiving messages pertaining
that sensitive case exchanged within the remaining members of the
department.

Beside information flow, ASNs require a way to manage the privileges
of their members.  In the following we define as privileges the
operations that a user is allowed to perform in a ASN.  To do that
\prism{} uses an approach similar to
\cite{DBLP:conf/sacmat/GeorgiadisMPT01}.  \prism{} uses the
\emph{roles} assigned to user by the ASN administrator.  In the
presented model a role is a job function/title within the organization
with some associated semantics regarding the authority and
responsibility conferred on a member \remove{of the} role.

\begin{definition}[Role]
\label{def:role}
Given an autonomous social network $asn \in \mathcal{ASN}$, a role $r
\in \mathcal{R}(asn)$ is a couple of the form $$\left< n,
  \mathcal{PR} 
\right>$$ where $n$ is the (unique) identifier of the role and
$\mathcal{PR}$ is the set of privileges granted and/or denied to the
members of the role $r$.
\end{definition}

As for groups, role may be organized hierarchically.

\begin{definition}[Role hierarchy]
\label{def:role_hierarchy}
Let $\mathcal{R}$ be a set of roles defined for an autonomous social
network $asn$. A role hierarchy $$\phi_\mathcal{R} : \mathcal{R}
\rightarrow \mathcal{R} \cup \{\bot\}$$ is the function which defines
the child-parent relationship among the roles.
\end{definition}

A user may be associated with multiple roles, according to the
functions she/he is performing within the organization.  Furthermore,
\prism{} allows the administrator to further refine the privileges
available to a given user according to ``where'' (in which context)
she/he is operating.  In fact the privileges granted to a given user
at a given moment are defined combining the roles to which the user
has been assigned and the subdomain in which she/he is operating.
Thus, the subdomains contribute to identify the available privileges,
refining the privileges of a role (both granting or revoking
privileges) or even granting/revoking permissions directly to specific
users.

Beside that, a group creator may wish to restrict the membership to
the group, for example not granting the membership to those users who
are member of another specific group.  Moreover, one may need to
moderate the messages associated with a given group.  \prism{}
provides to the users \remove{with the right privileges} the
possibility to specify \emph{group privileges} (to be elaborated in
Section~\ref{sec:privileges}).

Finally, we have all the concepts required to formally define circles.
\begin{definition}[Circle]
\label{def:circle}
Given a social mesh $SM$, the set of circles $\mathcal{C}$ 
 is defined as follows:
$$
\mathcal{C} = 
\mathcal{PG}(SM)
\bigcup_{asn \in \mathcal{ASN}(SM)} \mathcal{SD}(asn) 
\bigcup_{u \in \mathcal{U}(SM)}\mathcal{PRG}(u)
$$
\end{definition}

To wrap up, we define ``a message'', which is the entity of data created by and shared among the users of the social mesh.

\begin{definition}[Message]
\label{def:message}
A message $m$ is a tuple $\left<u, t, \mathcal{T}, \mathcal{I} \right>$ where $u \in \mathcal{U}$ is the author of the message, $t$ is the content.
$\mathcal{T}, \mathcal{I} \subseteq \mathcal{C}(u)$ are respectively called the tag and the conflict set.
\end{definition}


\section{Frontier Information Propagation Mechanism}
\label{sec:ac}

It is fairly complex to manage the communication within large
  organizations. In particular, sometimes it is not completely clear
  who are the users entitled to access certain information. The
  complexity increases rapidly when dealing with the
  communication between users belonging to different organizations. In
  the following, we will present a mechanism to handle such
  complexity, by taking advantage of the model defined in
  Section~\ref{sec:model}.
In our model, information propagation is performed with respect to
circles but not to domains since domains deal with privileges of
users, while circles have been specifically designed to deal with
information flow.

 The Frontier Information Propagation Mechanism ensures that a given message $m$ is accessible
by all the users who are member of at least a circle in
$\mathcal{T}(m)$ but who are not member of any circle in
$\mathcal{I}(m)$. In addition, other users may read the message $m$ 
satisfying

 the policies of at least a circle $c
\in \mathcal{T}(m)$.
 Moreover, it is also possible for a user to
access $m$ if there exists a sequence of circles $CSeq = c_1,\ldots, c_n$
where $c_n \in \mathcal{T}(m)$ and $\forall i \in [2, n], \phi(c_i) =
c_{i-1}$.
The user $u$ is allowed to access $m$ if and only if she/he satisfies the policies defined for all the circles in $CSeq$.

The syntax to describe the policies is outside the scope of this work and treated in works such as \cite{Weitzner05creatingthe}.

Informally, policies are of the form:
$$ a \leftarrow pred_1 \wedge \ldots \wedge pred_k $$ where $a \in \left\{\texttt{allow}, \texttt{deny}\right\}$ and each predicate $pred_i$ verifies properties of the message, the author of the message or the user reading the message.
The properties verified by the predicates currently supported by \prism comprehend: author/reader identity, author/reader membership, tags of the message, etc.

The enforcement mechanism is described in Alg.~\ref{algo:fipm}.

\begin{algorithm}[htb]
  \KwIn{$m$, the message to be accessed, $u$ the reader of the message}
  \SetKwFunction{verifies}{verifies}

  \Begin{
      \For{$c \in \mathcal{I}(m)$}{
       \If{ $u \in \mathcal{M}(c)$}{
         \Return \texttt{deny}\;
       }
      }

      \For{$c \in \mathcal{T}(m)$}{
        $c' := c$\;
        \While{$c' \neq \bot$}{
          \eIf{ $u \in \mathcal{M}(c')$}{
            \Return \texttt{allow}\;
          }{
            \If{\verifies{$u$, $\mathcal{P}(c')$}}{
              $c' := \phi(c')$\;
            }
          }
        }
      }

      \Return \texttt{deny}\;

     }

    \caption{The Frontier Information Propagation Mechanism.}
    \label{algo:fipm}
\end{algorithm}

\begin{figure}[htb]
\centering
\includegraphics[width=.35\columnwidth]{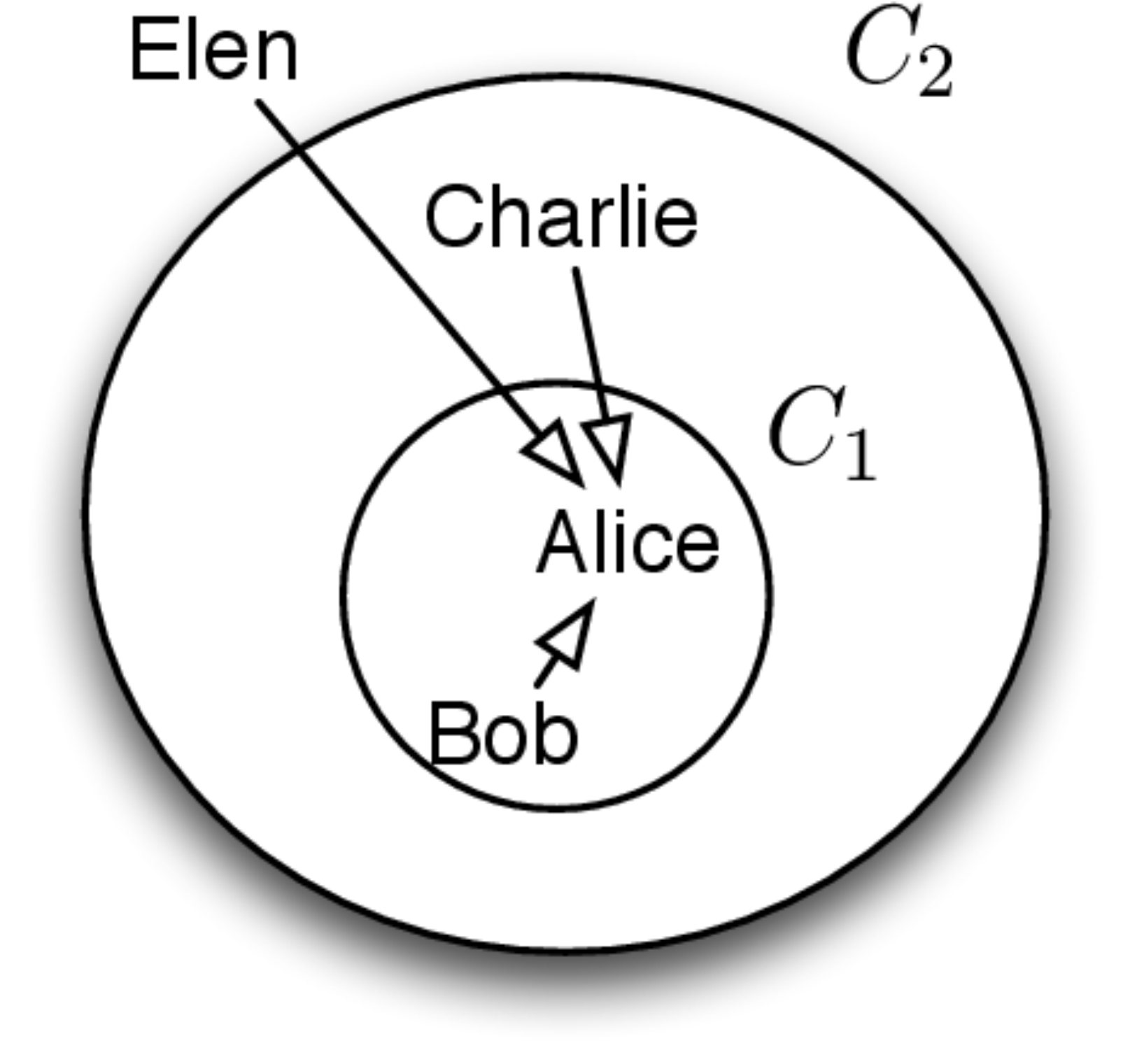}

\caption{An example for the frontier information propagation mechanism.}
\label{fig:circles_ac}
\end{figure}

Consider an example scenario shown in Figure~\ref{fig:circles_ac}.  In
such a scenario, the users Bob, Charlie and Ellen are following Alice.
Alice is member of the circle $C_1$ which is in turn an inner circle
of $C_2$.  Suppose Alice creates a message $m$ such that
$\mathcal{T}(m) = \left\{C_1\right\}$ and that $\mathcal{I}(m) =
\emptyset$.  As previously defined, the Frontier Information
Propagation mechanism states that if $\exists c \in \mathcal{T}(m)$
such that $reader \in \mathcal{M}(c)$ then $reader$ is allowed to
access the message.  Thus Bob is allowed to access $m$ since he is a
member of $C_1$.  On the other hand the other users will satisfy the
policies of $C_1$ to access $m$.  Assuming that both Charlie and Elen
satisfy such policies, only Charlie will access $m$ because he is a
member of $C_2$.  Hence, Elen will be required to satisfy also the
policies of $C_2$ before being able to read content from the circle
$C_2$.

If the collision set $\mathcal{I}(m)$ is not empty, then it needs to
be verified whether the reader is member of any of the circles in such
a set. If this is the case, then the reader is not allowed to access
$m$.

\section{Management of Privileges}
\label{sec:privileges}

\prism{} supports what we call \emph{group} and \emph{domain} privileges.
The former are those privileges which define the actions users can perform within a group, such as the privileges of joining the group, to tag a message with the current group or the requirement of the messages tagged with a group to be moderated by a boss of the group.
The latter are those privileges granting to users administrative powers, such as the privileges to create public circles, to create subdomains, to create roles and so on and so forth.

Group privileges are specific for the group
\emph{in which they are defined} and therefore their
enforcement is straightforward: once a user is operating in a specific
group, the group privileges are enforced.

In contrast, domain privileges require a more complex mechanism to be
enforced.  Note that the \prism{} framework manages and enforces
access control at ASN's level, in the sense that the domain privileges
are defined in groups characteristics of a ASN -- such are roles and
subdomains -- and they can be enforced only within the specific ASN.

As presented in Section~\ref{sec:model}, the operations a user is
granted to perform are defined by a combination of her/his roles and
the subdomain in which she/he is operating.  Because of that, the
\prism{} framework enforces access control differently according to
the action performed by the user.

\begin{figure}[htb]
\centering
\includegraphics[width=.6\columnwidth]{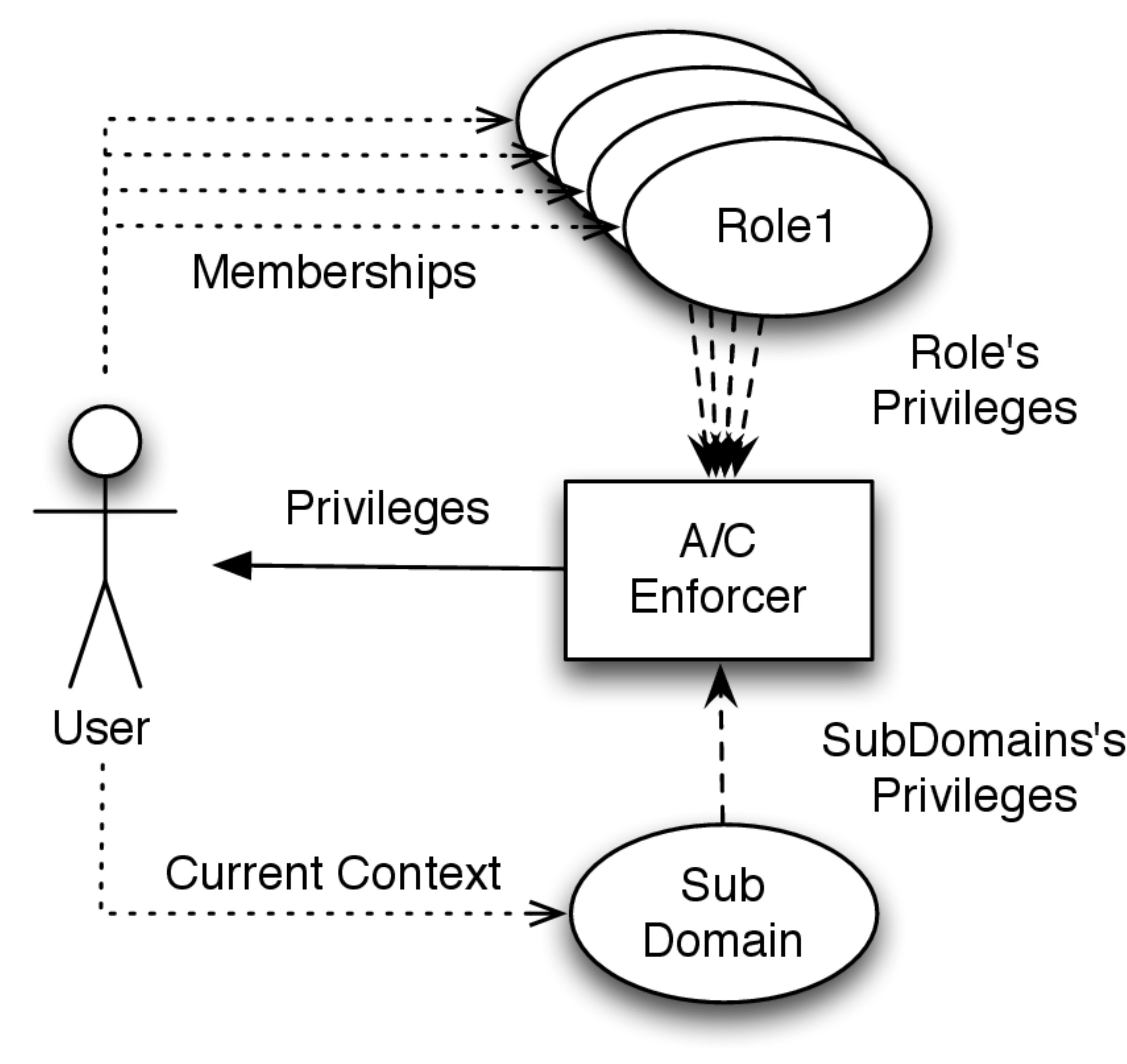}
\caption{Access control model.}
\label{fig:ac_model}
\end{figure}

The enforcement algorithm works as follows (also see
Figure~\ref{fig:ac_model}). Let us assume a given ASN
$asn$ and the user $u \in \mathcal{U}(asn)$ who is associated with the
roles $r_1,\ldots,r_n \in \mathcal{R}(asd)$. Thus, $u$ is granted the
privileges $u_{\mathcal{PR}} = \bigcup_{i=1}^n \mathcal{PR}(r_i)$.
When $u$ operates within a subdomain $sd \in \mathcal{SD}(asn)$ the
privileges actually granted to $u$ are computed as:
$$u_{\mathcal{PR}} \otimes \mathcal{PR}(sd)$$

Recall that the predicate $\otimes$ refines the privileges in $u_{\mathcal{PR}}$
with the ones defined in $\mathcal{PR}(sd)$.

\section{\prism{} architecture}
\label{sec:architecture}

\begin{figure}[htb]
\begin{center}
\includegraphics[width=.7\columnwidth]{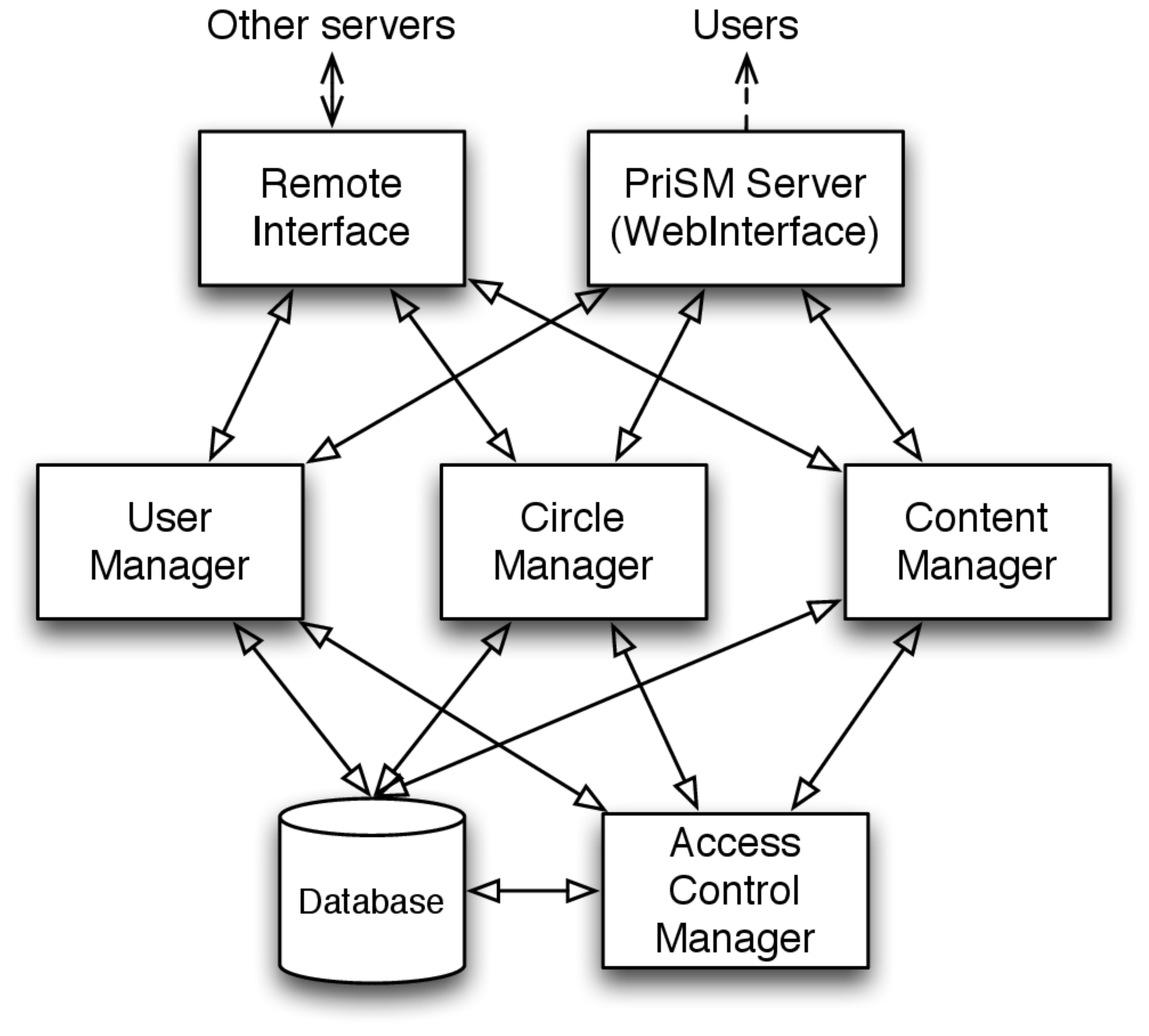}

\caption{PriSM ASN architecture.}
\label{fig:architecture}
\vspace{-2ex}
\end{center}
\end{figure}

In order to provide the services required by an ASN, each domain
deploys \prism{} locally. Figure~\ref{fig:architecture} shows the
architecture of an independent ASN deployment comprising several
interconnected modules. Each module is in charge of managing a
specific subset of the features provided by the system. Many of these
features are `standard' in any state-of-the-art online social network
platform while a few others are novel, specific to \prism{}'s
distributed/federated nature and its access and information flow
controls:

\begin{itemize}
\item
 \textbf{User Manager}: This module provides an interface to the operations directly related to the users, such as registration, profile management, relations and subscription of messages from other users, etc.

\item \textbf{Circle Manager}: This component controls the circles
  related information such as the lists of members and the propagation
  policies for each circle other than any relationships between them
  (See
  Definition~\ref{def:subdomain_hierarchy} 
  and Definition~\ref{def:collection_hierarchy}).

\item
 \textbf{Access Control Manager}: This module regulates both the actions performed by the users of a \prism{} ASN with respect to the privileges assigned to them by the domains administrators and enforces the policies defined in the circles (the later is elaborated in Section~\ref{sec:message_propagation}).

The functionalities of this module are: (i) to store and propagate the messages (and content) generated by the ASN's users and (ii) 
to grant access only to those users who are allowed according to the rules.
\end{itemize}
The \prism{}
 Web Interface exposes the services orchestrated by all these constituent modules to the ASN users.

 A final module manages the interconnections between the different ASN
 instances of \prism{}.

\begin{itemize}
\item
  \textbf{Remote Interface:} This module is in charge of performing
  the operations of exchanging information with other ASNs. For
  example, the Remote Interface retrieves the required data when a
  user is accessing the profile of some user $u'$ in some other domain
  $D'$. It also sends to the interested domains the updates involving
  shared data, such as those regarding the members and/or the policies
  of shared circles.

  The present \prism{} implementation allows communication between
  only ASNs which have been manually paired by the domains'
  administrators. Paired ASNs are considered trusted in the current
  model. Additionally, at present we assume the existence of a service
  to correctly discover other ASNs and their trustworthiness. These
  assumptions need further consideration in future. We will also like
  to \remove{note} remark that individual ASN deployments are free to
  tweak the constituent modules, to add or modify functionalities as
  deemed appropriate.

\end{itemize}

\subsection{Message propagation}
\label{sec:message_propagation}
The primary objective of the \prism{} system is to allow users to
exchange information. In order to provide the users a
satisfactory experience, the architecture of \prism{} has been designed
to reduce the time elapsing between when the
information is created and when it is actually available to the final
user.

\begin{figure}[htb]
\centering
\includegraphics[width=.7\columnwidth]{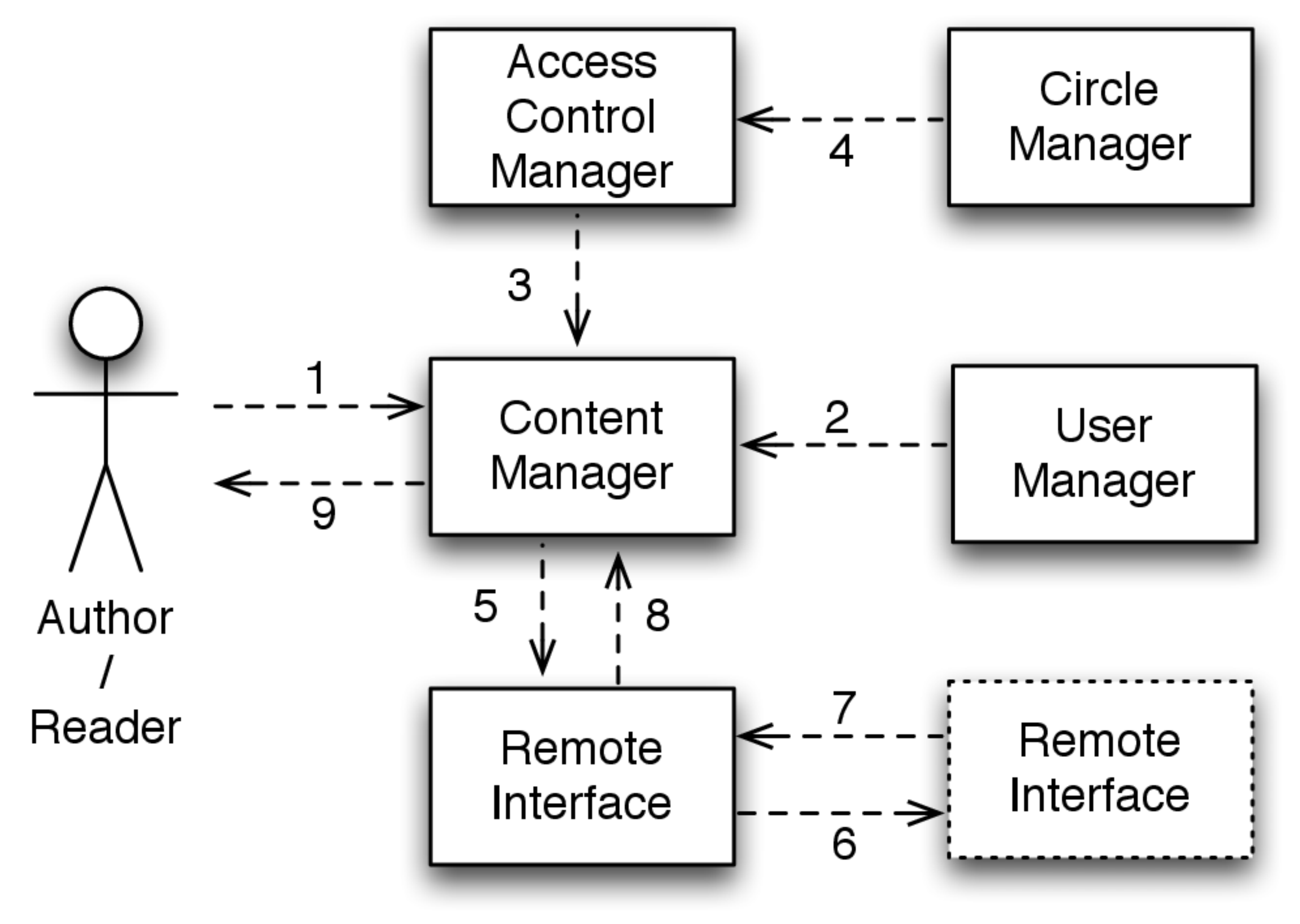}

\caption{How to post/retrieve a message in \prism{}.}
\label{fig:posting_procedure}
\end{figure}

Figure~\ref{fig:posting_procedure} shows the steps required to post a
message through the system to all the users potentially interested in
\remove{such message which stated to be interested into the author}
it.  First of all the user $u$ sends the message $m$ to the Content
Manager (1), which stores the message in the local database.
Afterwards, the Content Manager retrieves the set of followers
$\hat{\mathcal{F}} = \left\{u_1,\ldots, u_k\right\}$ from the User
Manager (2).  The Content Manager requests to the Access Control
Manager for each local user $u_i \in
\mathcal{L} 
\subseteq \hat{\mathcal{F}}$, if $u_i $ is allowed to access $m$ (3).
The verification is performed by Access Control Manager according to
both the tag set, the conflict set (see Definition~\ref{def:message}),
the set of circles to which $u'_i$ is member of and the list of
propagation polices. Such information are retrieved by the Access
Control Manager querying the Circle Manager (4).  If the verification
(3) holds then the Content Manager will notify the user $u_i$,
immediately if the user is currently online or delivered in the user's
`inbox' to be retrieved as soon as she/he logs into the system (9).
At the same time, the Content Manager sends the set of remote users
$\mathcal{RU} = \hat{\mathcal{F}} \setminus \mathcal{L}$ to the Remote
Interface (6) which will, in turn, extract the set of domains
$\mathcal{RD} = \left\{d_1, \ldots, d_q\right\}$, with $q \leq \left|
  \mathcal{RU} \right|$, to be notified of the existence of $m$ (6).

\remove{Note that the notification to the remote domains is performed sending $m$.}
The action of notifying the remote domains actually consists in forwarding $m$.
Therefore, each remote domain $rd \in \mathcal{RD}$ will send the message $m$ to the \emph{local} Content Manager (8) which, in turn, will perform the steps (2) to (4), as performed by the Content Manager of the original domain, including the final notification (9) to the users local to $rd$.

We assume each domain to be trusted. It means that the Access Control
Manager will behave consistently across all ASNs.  Moreover, we assume
that circles' data and messages will be replicated among different
domains, mainly to reduce the latency of the system.  Note that such
an assumptions do not introduce any vulnerability substantially
different than while using other modes of electronic communication
such as email.


\subsection{Implementation of the framework}
\label{sec:framework}

The model described in Section~\ref{sec:model} and the architecture
previously presented have been implemented in a prototype using Java 6
and GWT~\cite{GWT} for the user interface.  A MySQL~\cite{MySQL}
database is \remove{been} used for the persistency of the data.  The
communication protocol between the different deployed ASN instances
occurs using a well defined REST
interface~\cite{DBLP:www/org/w3/http1-1}.

The current prototype is structured as a modular server, in which each component is directly connected with the others as shown in Figure~\ref{fig:architecture}.
Nevertheless, the server modules can be easily separated on different machines, to take advantage of such parallelism.

A final remark we will like to make, to repeat what has been stated
elsewhere, is that the individual modules can be modified, or
additional modules added, as deemed essential for an ASN
instance. Furthermore, we are working on exposing a set of interfaces
so that other ``apps'' can be deployed on top of \prism{} by
leveraging on its existing functionalities.

\subsection{Evaluation and discussion of the architecture}
\label{sec:evaluation}

From our observations, the \prism{} architecture presents mainly three
possible scalability bottlenecks (i) the Web Interface (ii) the Remote
Interface and (iii) the database.  More precisely, increasing the
number of the users of a ASN increases the probability of users
connected simultaneously to the system.  Such a condition will require
an ever increasing amount of computational resources.  Similarly, more
resources are required to provide the same promptness of the system
with the increase of the number of interconnected ASNs.

The three previously mentioned issues can be addressed using standard
distributed systems techniques, such as replicating the appropriate
modules of the architecture.

Each module of the \prism{}'s architecture is stateless,\footnote{See
  \cite{Fielding:200:AS} for a more formal and complete definition of
  stateless} and internally highly parallelized, specifically with the
intent to simplify its replication.  Similarly, the database can be
replicated and distributed as well.  However, such operation will have
a cost in terms of an increased complexity to manage the consistency
of the data.

\begin{figure}[htb]
  \centering
  \centering
\includegraphics[width=.95\columnwidth]{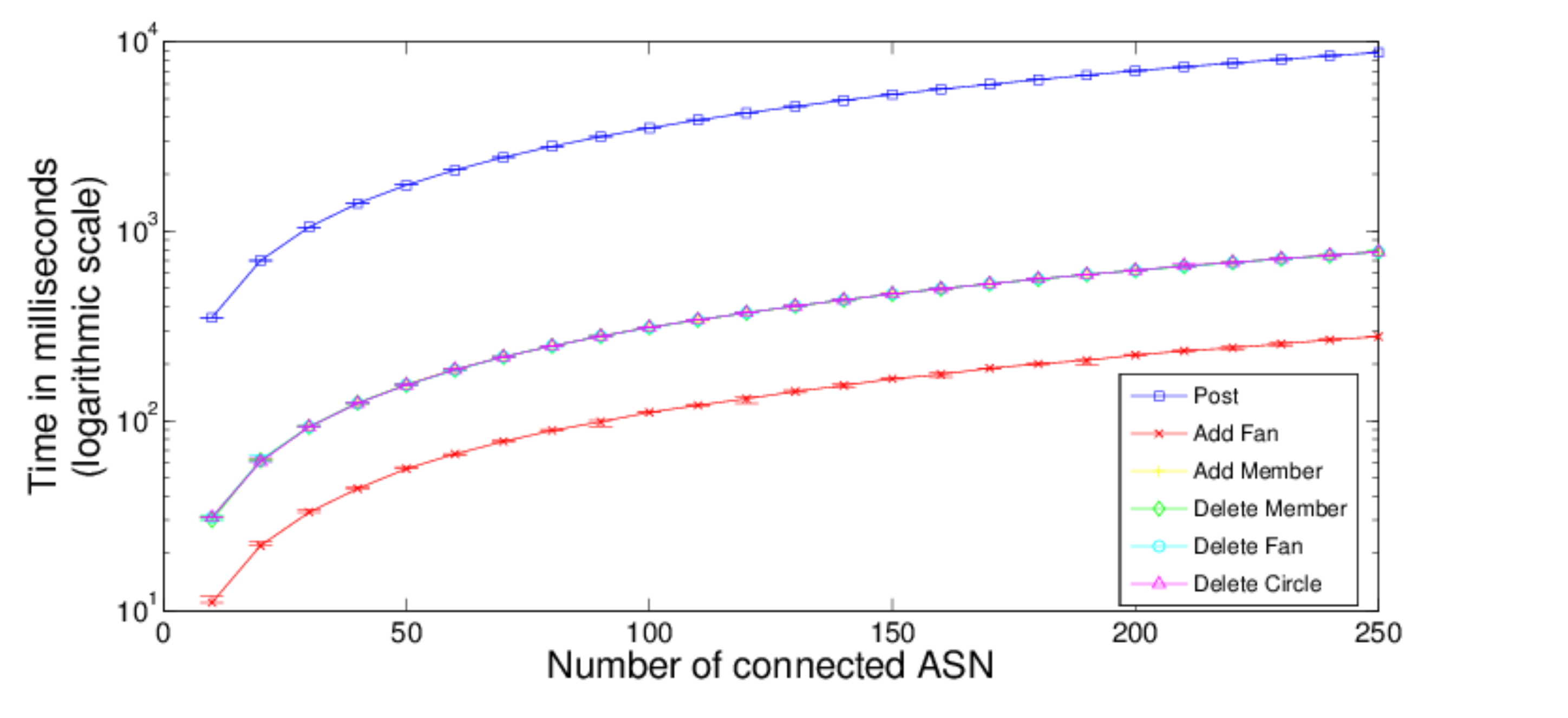}

  \caption{Empirical evaluation of the architecture's scalability, the y-axis uses a logarithmic scale.}
  \label{fig:scalability}
\end{figure}

We benchmarked the performances of the remote operations to
empirically verify the scalability of the proposed architecture.  We
evaluated the execution time of each remote operation varying the
number of involved ASNs.  The experiments have been executed in a
network of two computers (Linux 3.0.1 running on a Intel Core 2 Duo
2.53GHz with 4GB of RAM).  On the first machine we ran the \prism{}
prototype, while on the other machine ran a `light weight' version,
which did not provide the Web Interface.  The results are shown in
Figure~\ref{fig:scalability}.  As one may notice, the time required
for the execution of each operation is negligible except for sending
messages to the remote ASNs -- the Post operation.  We observed that
\prism{} prototype requires on an average 8780 milliseconds to
propagate a message to 250 distinct ASNs. Note that, normally an
individual has 100s of contacts. Thus, even if each of these contacts
were to belong to a different ASN, the delays introduced for scaling
the message propagation over very many ASNS is reasonable, showcasing
the scalability of our \prism{} implementation.

\begin{figure}[htb]
\centering
\includegraphics[width=.95\columnwidth]{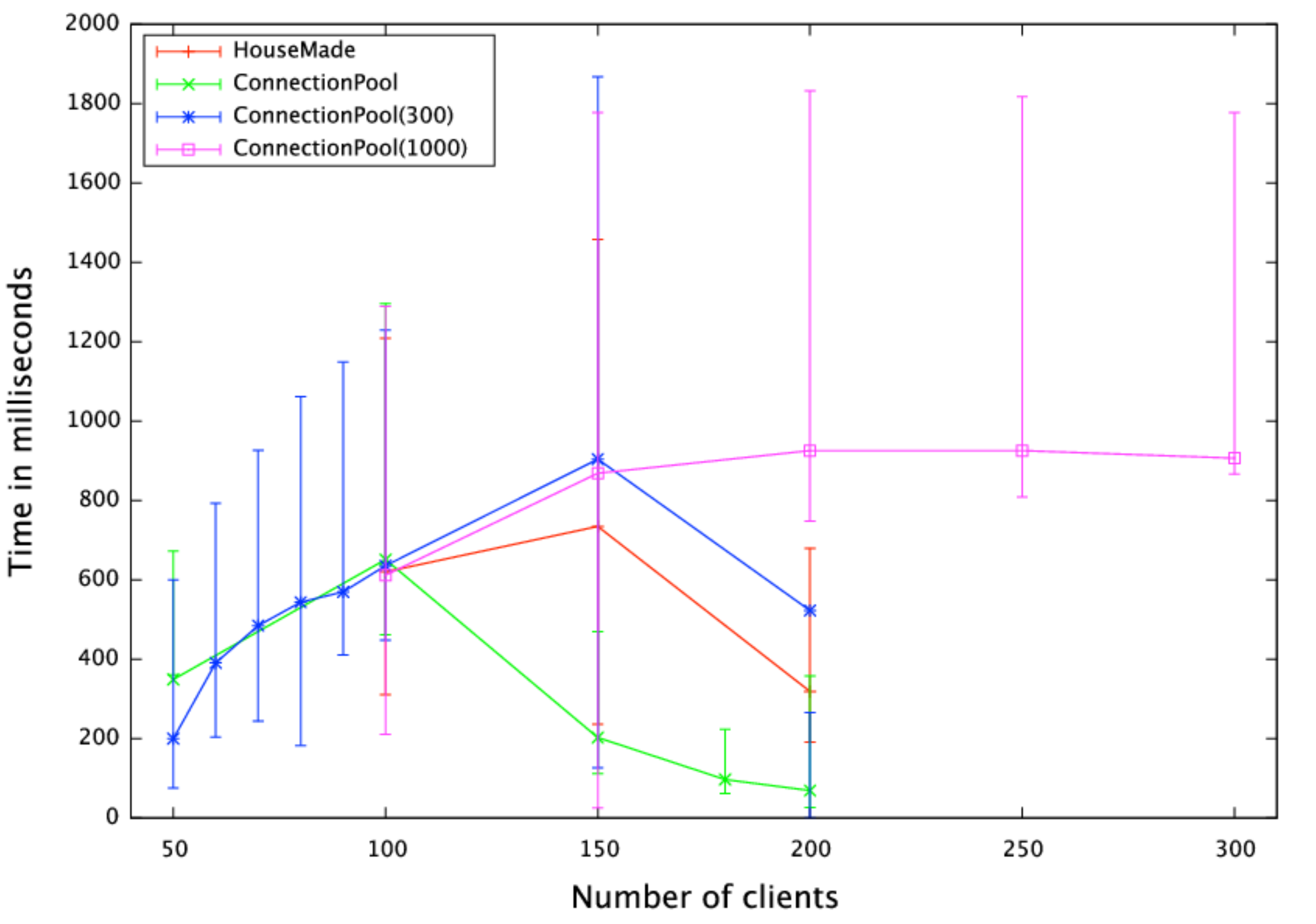}
\caption{Number of varying the number of requesting users, with the 10th and 90th percentile.}
\label{fig:capacity}
\end{figure}

Given the result of the previous experiments, we also evaluated the
scalability of the software itself. This was done by evaluating the
capacity level of a single PriSM instance. Thus, we measured the
number of operations per second that a PriSM instance is able to
handle under different stress levels.  In this experiment we executed
only the \textbf{Post} operations because, as we previously showed, it
is the most expensive in term of resources required. To create
different load levels we set as a parameter the number of parallel
clients (from 50 to 300).  Figure~\ref{fig:capacity} shows the
results. Our first implementation was not really able to scale that
well (see the \emph{HouseMade} series in
Figure~\ref{fig:capacity}). In fact we obtained several
database-related error messages when we hit the prototype with 150
clients. Initially we imagined that the errors were caused by bad
performances of our house-made connection pool. Hence, we changed to
the one provided by Apache Tomcat.  The results using the default
configuration were even worse (see the \emph{ConnectionPool}
series). We changed then the configuration increasing the minimum
number of pooled connections (from 100 to 300) and the results
increased (\emph{ConnectionPool(300)} in the graph) but not as much as
we expected. Finally, we realized that the bottleneck was caused by
the configuration of the database server instead of the connection
pool.  This was caused mainly by the fact that the messages generated
by MySQL are fairly cryptic.

Hence, we extended the log mechanism of \prism{} to better observe the
usage of the connection pool and of the various connections. The
primary outcome was a better understanding of the errors. MySQL is by
default configured to handle 100 parallel connections through the
network. Increasing such value to 1000 partially improved the
performances because the machine's operating system (Linux Ubuntu
12.04) has a fairly long \texttt{TIME-WAIT} for actually closing
connections (see \cite{RFC0791}).

Once we modified the corresponding configuration, both increasing the
number of parallel connection and setting the configuration of the OS
to reuse connections in \texttt{TIME-WAIT} state, we were able to
measure the real capacity of the system (as shown in
Figure~\ref{fig:capacity}, \emph{ConnectionPool(1000)} ).

Furthermore, we evaluated the time taken to post a message to
different ASNs.  To do that, we took advantage of a real cloud
provider. Namely, we rented 5 EC2\cite{ec2} instances in 5 different
datacenters across the globe (see Table~\ref{table:datacenters}).  All
the instances have the same (virtual) hardware configuration:
\textbf{m1.large}, 7.5 GiB memory, 4 EC2 Compute Units (2 virtual
cores with 2 EC2 Compute Units each), 850 GB instance storage, 64-bit
platform.  We configured the virtual machines to run Ubuntu 12.04 LTS
and we installed on them the strictly required software (OpenJDK 7,
Apache Tomcat 7, MySQL 5.1).  The experiments run as follows: we
simulated that a user posed a message in the Singaporean ASN and we
measured the time required to the message to reach all the other ASNs.

\begin{table}[htb]
\centering
\begin{tabular}{| c | c | c |}
\hline
\textbf{Name} & \textbf{Location} & \textbf{Avg ping}\\
\hline
\hline
\verb|asia_sg| & Asia Pacific (Singapore) & 0ms\\
\verb|eu_west| & Europe (Ireland) & 550ms \\
\verb|s_america| & South America (Brazil) & 737ms\\
\verb|us_east| & US East (N. Virginia) & 526ms\\
\verb|us_west| & US West (Oregon) & 445ms\\
\hline
\end{tabular}
\caption{Characteristics of the EC2 instances.}
\label{table:datacenters}
\end{table}

The parameter for these series of experiment was the number of threads
handling the delivery of the messages.

\begin{figure*}[htb]
  \centering \subfloat[1 thread\label{fig:deliver1}]{\includegraphics[width=.85\columnwidth]{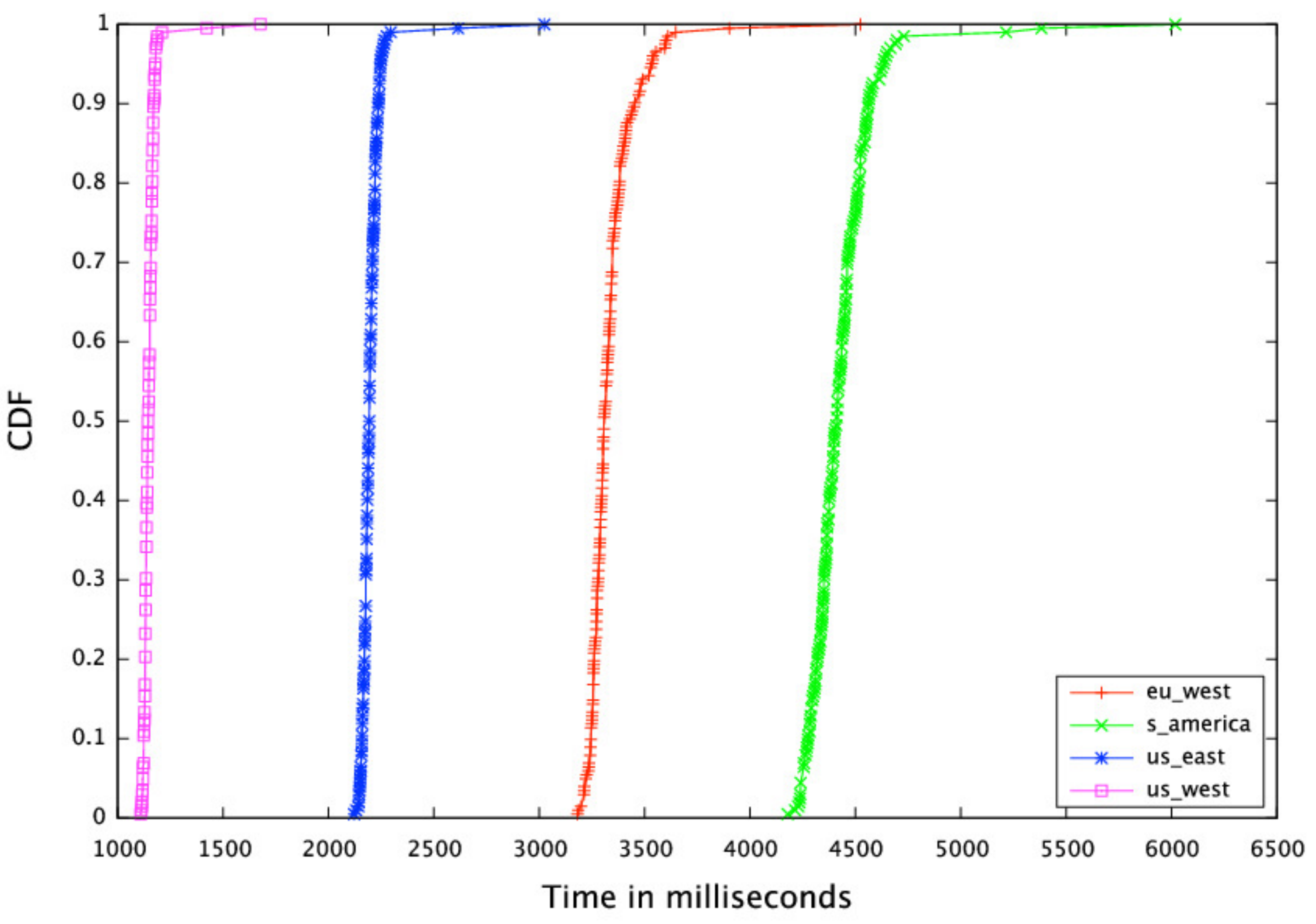}
  } \quad \subfloat[2 threads\label{fig:deliver2}]{\includegraphics[width=.85\columnwidth]{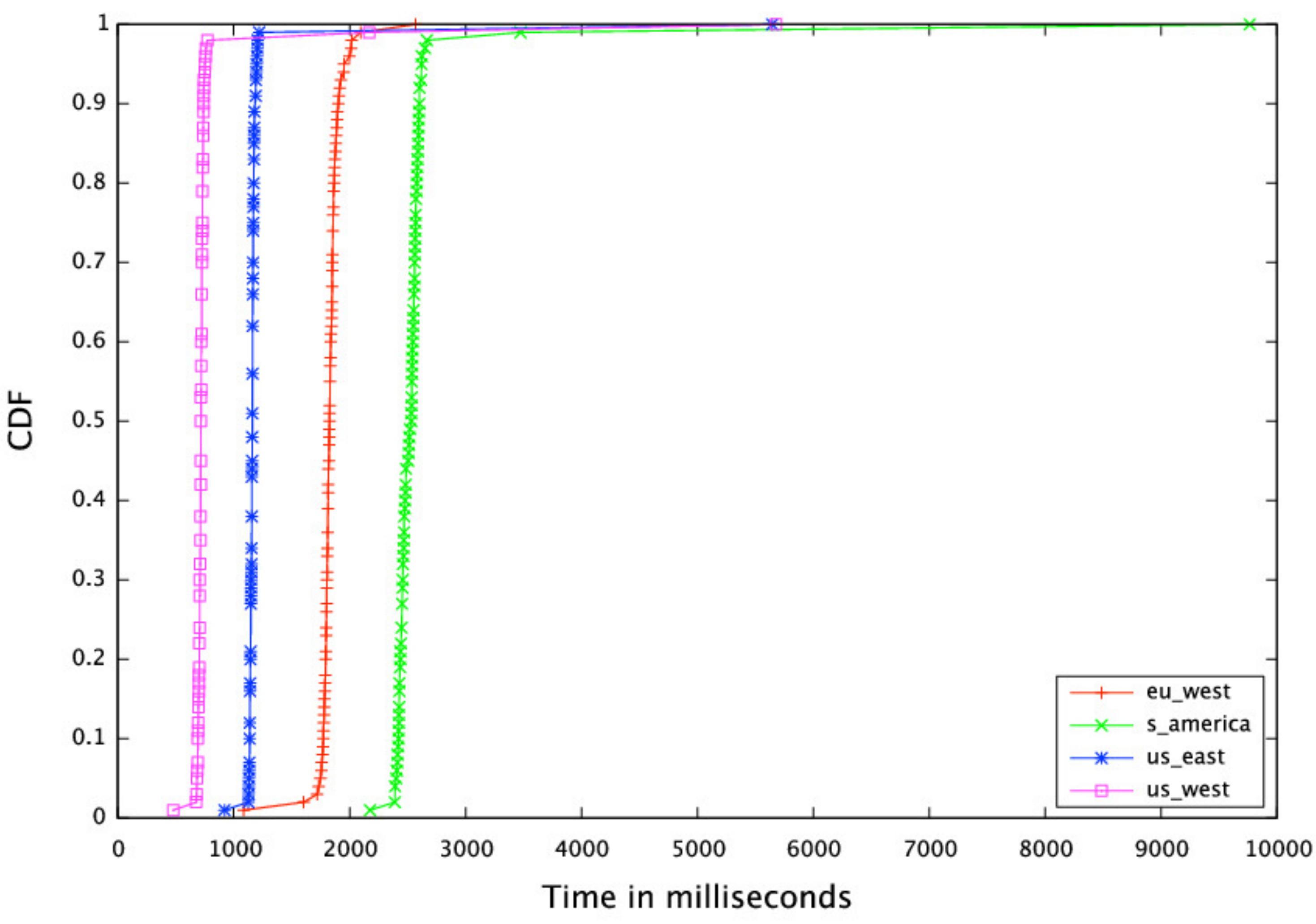}
  }\\
  \subfloat[3 threads\label{fig:deliver3}]{\includegraphics[width=.85\columnwidth]{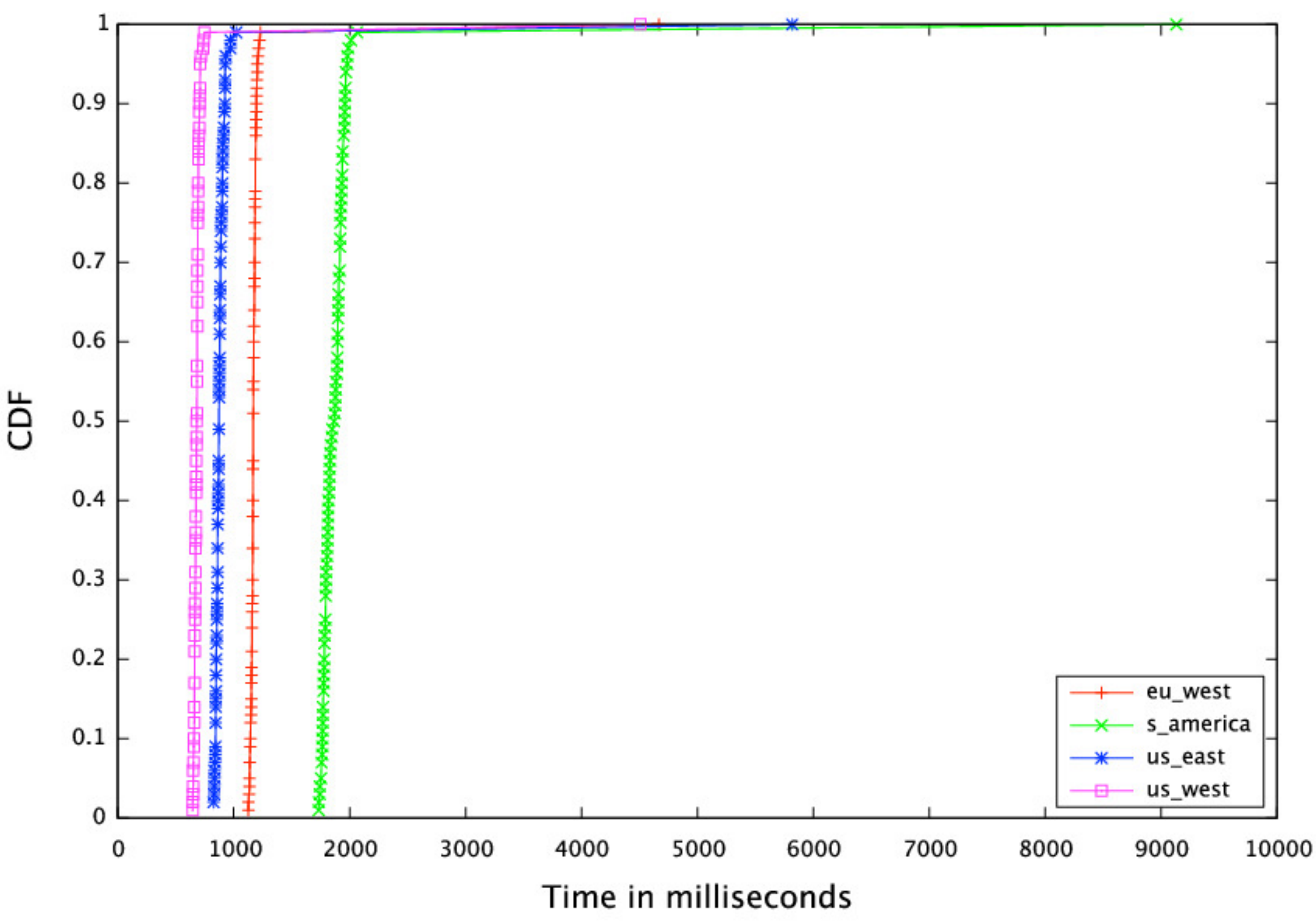}
  } \quad \subfloat[4 threads\label{fig:deliver4}]{\includegraphics[width=.85\columnwidth]{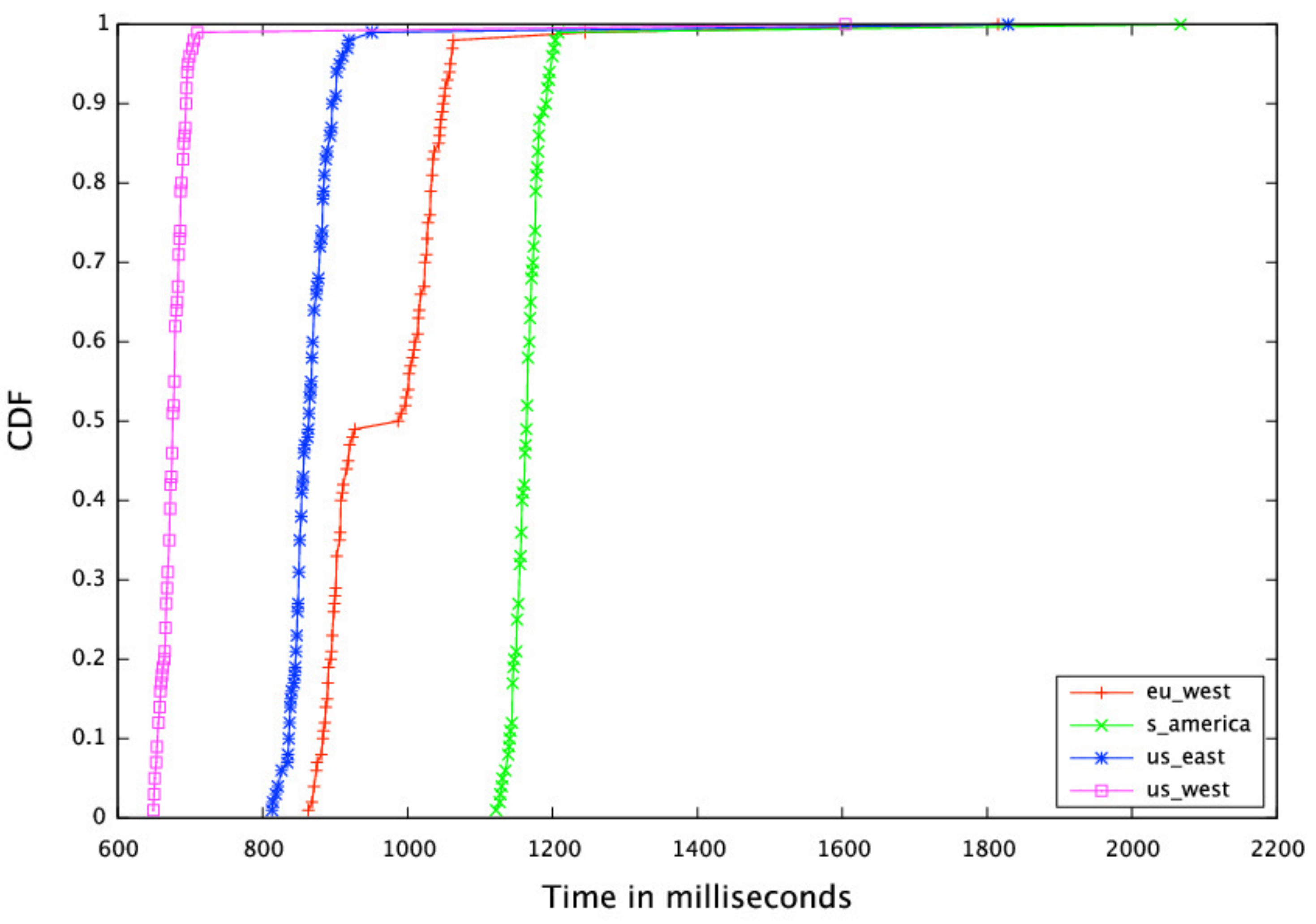}
  }
\caption{Remote post execution time distribution with different number of threads.}
\end{figure*}

The results of each experiment are shown in Figure~\ref{fig:deliver1},
Figure~\ref{fig:deliver2}, Figure~\ref{fig:deliver3} and
Figure~\ref{fig:deliver4} respectively. As we expected increasing the
number of working threads decreased the time required to deliver the
messages to remote sites.

\begin{figure}[htb]
  \centering
\includegraphics[width=.95\columnwidth]{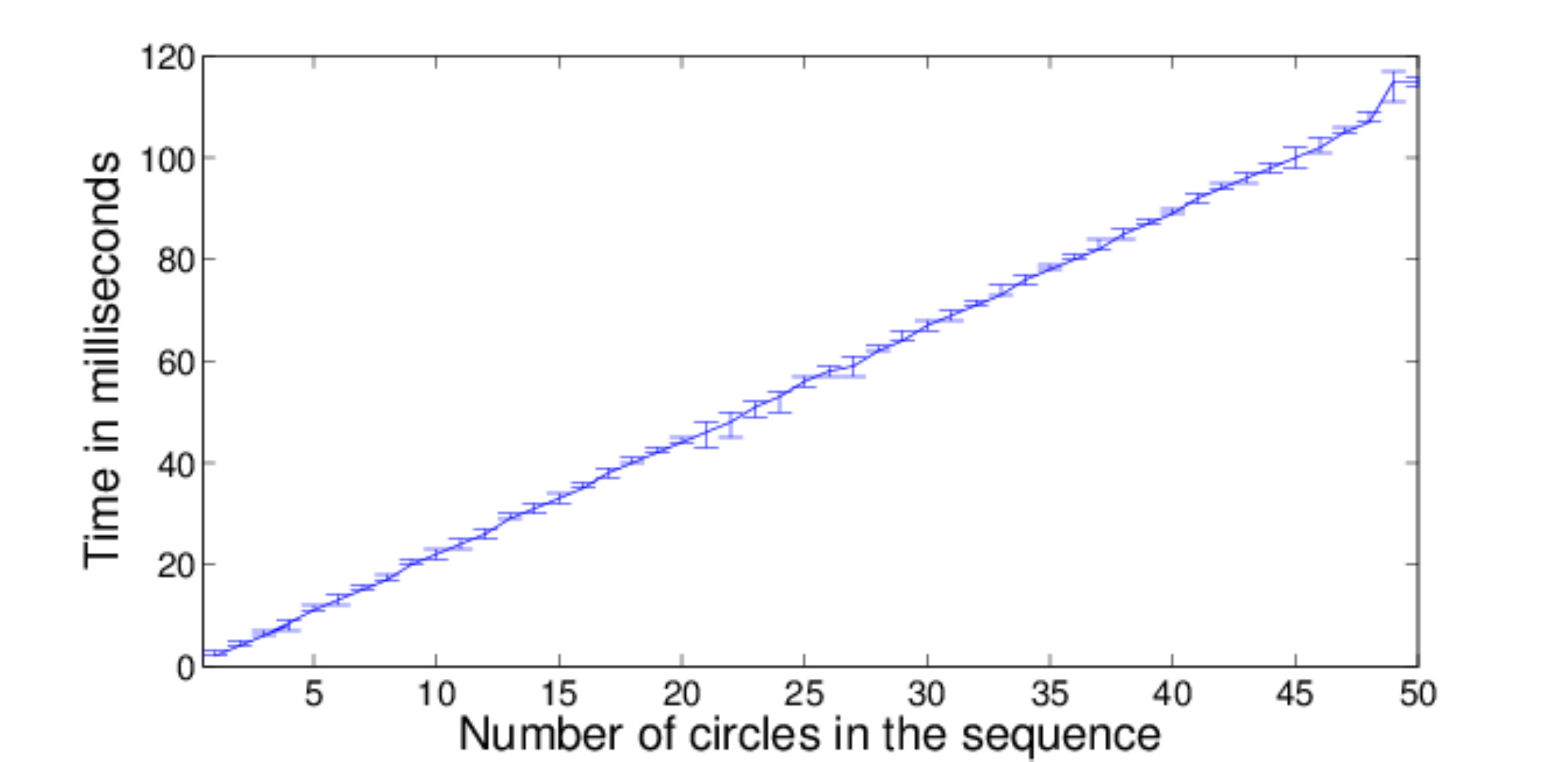}
  \caption{Time required to access a message with respect to 
   the number of circles in the sequence; the error bar represents the 10th and the 90th percentiles.}

\label{fig:acScalabilityA}
\end{figure}

We also benchmarked the time required for a user to access a given
message. In particular we evaluated two aspects: the length of the
sequence of circles that the reader has to cross in order to access
a message (see Section~\ref{sec:ac}) and the number of rules to be
evaluated to access messages of a given circle.  In both scenarios we
assume a ASN with 100 circles hierarchically organized and a user $u$
who wants to access a message $m$. $u$ is member of 50 random circles
and $\mathcal{T}(m)$ contains 10 (random) circles $c_1, \ldots,
c_{10}$ with $\forall i \in [1, 10], u \not \in \mathcal{M}(c_i)$.  We
also define 10 rules composed by 10 random predicates for each circle.
As shown in Figure~\ref{fig:acScalabilityA}, the time required for $u$
to access the message $m$ is linear to the length of the sequence of
circles separating $u$ from $m$.  We observed that on an average 115
milliseconds were required to access a message tagged with the last
circle of a sequence of 50 circles.  We also observed that the number
of tags associated with $m$ has a lesser impact on the performances.
The main reason is because the evaluation of the different possible
sequences of circles is performed in parallel and, more importantly,
distinct sequences are merged if during their evaluations common
circles are found.

\begin{figure}[htb]
\centering
\includegraphics[width=.95\columnwidth]{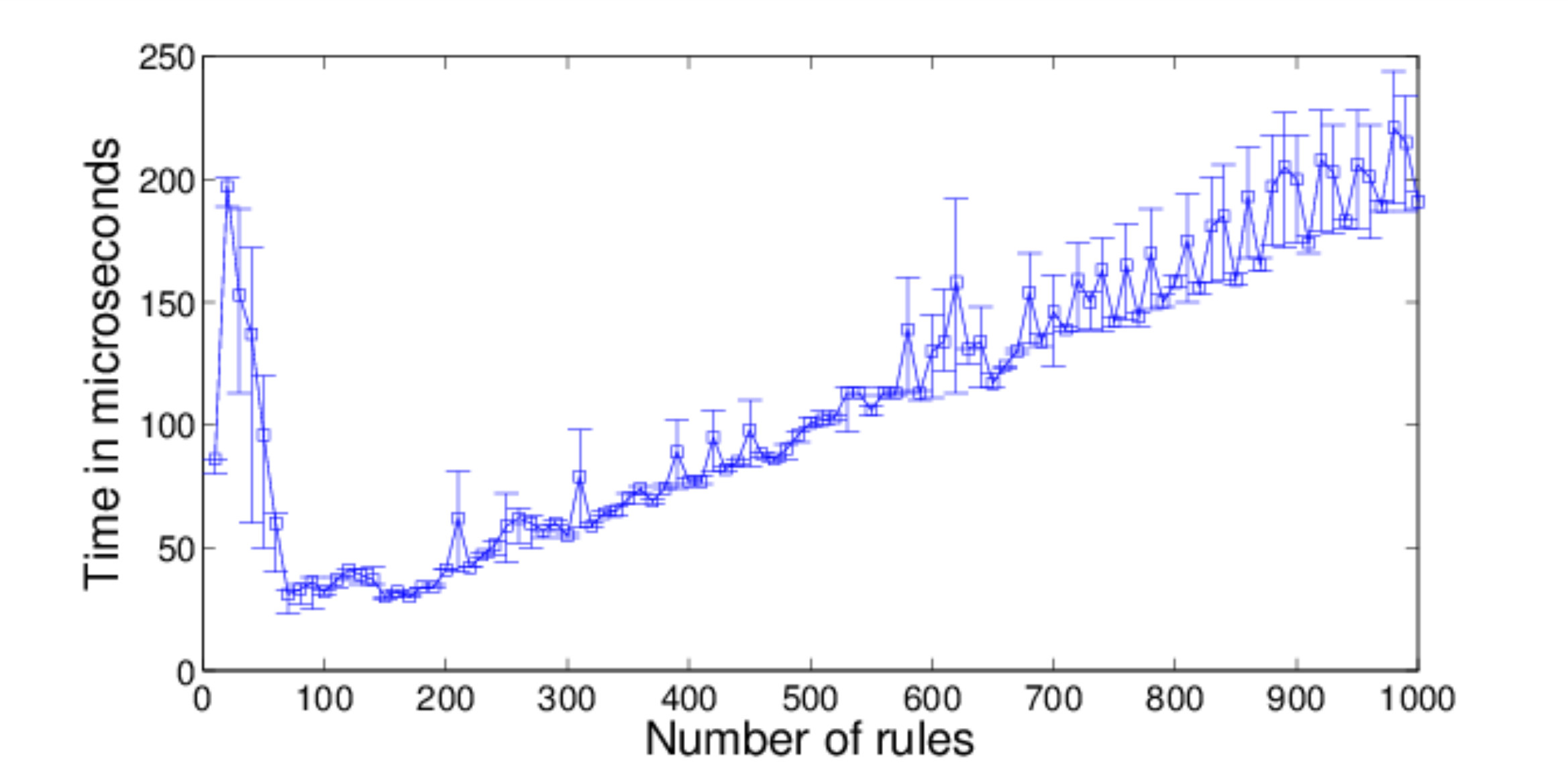}

\caption{Time required to access a message with respect to the number
  of rules; the error bar represents the 10th and the 90th
  percentiles.}
  \label{fig:acScalabilityB}
\end{figure}

In the second series of experiments, we investigate another aspect,
that of the time required by a user $u$ to access messages contained
in a given circle $c$, with $u \not \in \mathcal{M}(c)$ and $\phi(c) =
\bot$ (see Section~\ref{sec:model}).  As expected, the time required
is linear with the number of rules associated with $c$.  We observed
that on an average 187 microseconds were required to evaluate 1000
rules.  The results are shown in Figure~\ref{fig:acScalabilityB}.
Based on observations, the jitter trend in
Figure~\ref{fig:acScalabilityB} is caused by the memory allocation of
the JVM and noises from background processes running on the testing
machine.



\section{Related works}
\label{sec:related}
In the current section we briefly discuss some works related to the proposed framework.

\textbf{Decentralized Social Network.}
There has been recent interests in deploying decentralized online social networking (DOSN) as an alternative to the centralized third party services such as Facebook in order to avoid big brotherly controls and monitoring. Different architectures have been proposed by open-source as well as academic communities, which include Diaspora \cite{diaspora}, Appleseed \cite{appleseed}, Vis-{\`a}-Vis~\cite{DBLP:conf/comsnets/ShakimovLCCLLV11}, SuperNova \cite{supernova} among others \cite{p2p11-social}.
Traditional anonymous communication and file sharing networks such as Freenet \cite{freenet} have also been adapted to support friend-to-friend darknet subnetworks.
A more detailed survey on DOSNs can be found at \cite{DBLP:reference/social/DattaBVSR10}. The main motivation of these works is privacy, anonymity and free-speech of individuals. The deployment models are predominantly peer-to-peer in nature, where most (all) individual participants contribute resources to the system and control their individual data, and hence the infrastructure provides best effort service, and the design focus are towards dealing with system churn, fairness \& incentives, etc.

\textbf{Federated Social Network.}
Another criticism of centralized social networks like Facebook and LinkedIn is that users are tied-in, and cannot communicate across networks. Social network interoperability has been advocated to address such barriers~\cite{tbl,federatedWiki} for users to communicate across social networks and achieve portability.
In achieving federation of such social networks, the main challenge is to specify data format and protocol for exchanging information across different platforms - dealing with both technical issues (originating from the different existing implementations) and legal and commercial issues (e.g., companies are unwilling to expose user data to competitors).

\prism{} is instead designed for deployment in workplaces, and works
with similar assumptions of trust as do organizational email
services. Specifically, in \prism{} autonomous social networks (ASN)
are deployed and managed in a centralized manner on well provisioned
infrastructure. It gives its users privacy privileges with respect
to other fellow users, but not necessarily from the organization
whose infrastructure the users are using. Instead, it is designed to
provide the organizations a means to manage their users'
interactions flexibly and subject to the organizations' security and
confidentiality needs. The goals of federation are also distinct, in
that the federation is among multiple ASN instances with a common
set of communication interfaces, but the objective is to enable
flexible specification and control of information across ASNs, as
determined by organizational business logic and confidentiality
needs. Cross-platform federation of \prism{} ASNs with other social
networks will still need extrinsic mechanisms \cite{federatedWiki}.


\textbf{Commercial alternatives.} Oracle Social Network
\cite{oracle}, Yammer \cite{yammer} and SalesForce \cite{salesforce}
are a few commercial services providing some analogous
functionalities by facilitating inter-department and
inter-organization information exchange. These services reside in
third party cloud infrastructures, in contrast to \prism{}, which
owing to its decentralized architecture allows multiple deployment
models, including third party public cloud as well as fully
controlled private cloud hosting. Furthermore, \prism{} allows
customized mapping of organizational hierarchy \& workflows and
finer grained specification of who is entitled to access certain
information, a feature which is lacking in the existing general
purpose services.


\remove{Security and privacy preserving in social network have been an active research field.
More precisely, research in the field of access control has resulted
in the definition of some access control models and related mechanisms aiming to overcome the
restrictions of the protection mechanisms provided by current OSNs
 and \remove{As far as privacy is concerned,} current work is mainly focusing on protecting private information while performing social network analysis (see \cite{springerlink} and \cite{BFbook} for more details).}
\textbf{Security and Privacy.}
OSNs and DOSNs are often criticized for the currently provided protection mechanisms.
To overcome such restrictions several works has been done, mainly focusing on protecting private and sensible information while performing social network analysis (see \cite{springerlink, BFbook}).
One of the common characteristics of almost all the newly defined access control models is that access control is \emph{relationship-based}~\cite{Fong:2011:RAC:1943513.1943539}, that is,  authorized users  are denoted
on the basis of constraints on the  relationships the requester should have with other network users and/or the trust level associated with a relationship~\cite{DBLP:journals/tissec/CarminatiFP09}.
Following such trend, \prism{} natively supports an efficient relationship-based security mechanism based on relationships specified by means of circles, mimicing the security rules of work environments. Such mechanism can be easily extended to include more advanced constraints like the ones previously presented.

\textbf{Access control.} With respect to ``pure'' access control,
the most widespread family of access control model is RBAC
(Role-Based Access Control), proposed in
\cite{DBLP:journals/ac/Sandhu98} and in subsequent publications.
\prism{} provides management of privileges as in
\cite{DBLP:journals/ac/Sandhu98}, taking advantage of the role
concept with the objective to simplify the management of the
privileges assigned to users. Our approach is also inspired by the
works
\cite{DBLP:conf/rbac/Thomas97,DBLP:conf/sacmat/GeorgiadisMPT01}.
These works extend the RBAC model such that: the roles define the
actions that a user may perform while the ``team'' defines the
object on which such actions can be performed. In \prism{}, this
idea has been further extended using teams for refining the
privileges associated to roles. Practically, in \prism{} it is
possible for an administrator to create roles that are more general
than in a pure RBAC model, and therefore fewer number of roles
suffice. The context, defined by the team, can be used both to
identify the objects on which the user is allowed to operate and to
slightly change the privileges of the role. As a result, \prism{}
provides a way to reduce the increasing number of defined roles in
order to identify the right set of privileges for a given task. This
is a common and well known issue in systems deploying the RBAC
model. To balance that, \prism{} allows the operational context, the
team or in our case the subdomain, to grant/revoke privileges when
required.

\remove{The proposed access control model, on the other hand, doesn't consider at the current stage relationship or trust as parameter on which specifying constraints.
Nevertheless, it is possible to easily extend the current definition with such feature.
Moreover, the proposed model possess the advantage of defining deny rules, a feature which greatly simplifies the definition of complex rules.}


\section{Conclusion and future work}
\label{sec:conclusion}

In this paper we presented \prism{}, a framework for peer-to-peer
interactions among autonomous social networks. The \prism{} framework
is supported by a formal model defining relationships and interactions
among the different users. In particular the framework allows
delegated declaration/administration of (sub-)domains which allow the
possibility to define inherited privileges and restrictions on
individuals and groups of users, and provides easy to form
communication groups for the members to interact among themselves
subject to the constraints. While \prism{} facilitates confidentiality
and privacy aware communication across autonomous entities, thus
allowing organizations to retain ownership of data and control the
flow of information, it does not provide confidentiality to individual
users from the organization to which an user belongs.  Additional
cryptographic techniques would be necessary for the same. The modular
architecture of our implementation will allow such solutions to be
plugged in.

The proposed framework (and the prototype implementation) provides a
flexible solution for the deployment of collaborative network in
different application scenarios. These include (1) health sector,
where different kinds of entities and interactions are involved -
such as internal communication within and across hospitals, supply
chain management with pharmaceutical companies as well as public
relations, outreach and patient support groups, (2) customer
relationship management and enterprise resource planning for private
and public companies allowing collaboration for the fulfillment of
joint operations but still ensuring that the exchanged information
abide management policies, (3) educational environment complementing
existing e-learning tools for better intra/inter-institute
communication, (4) local/city-level administration, etc. We are at
the moment engaged in exploratory discussions with stake-holders
from several of these application scenarios to customize and deploy
\prism{} instances.

\remove{Advances in cloud operating systems (such as Mirage
\cite{mirage} allow developers to write network applications which
can be efficiently executed in the cloud environment directly as
virtual machines. Therefore, we are exploring ways to refine and
improve the current \prism{} implementation for a more portable
deployment.}

As part of future work, we aim to define a \emph{user API} allowing
the deploying
 organizations to create personalized extensions to the framework,
 taking advantage of all the features of the communication
 infrastructure.

Finally, we also intend to formally define and verify the frontier
information propagation mechanism, with respect to both the policy
definition language and the corresponding enforcing protocols.


\section*{Acknowledgement} The work presented in this paper was
partially supported by A*Star TSRP grant number 102 158 0038 and
NTU/MoE Tier-1 grant number RG 29/09.  S. Braghin did this work while
a research fellow at NTU Singapore.

\bibliographystyle{IEEEtran}
\bibliography{PriSM-unknown}

\begin{thebibliography}{10}
\providecommand{\url}[1]{#1}
\csname url@samestyle\endcsname
\providecommand{\newblock}{\relax}
\providecommand{\bibinfo}[2]{#2}
\providecommand{\BIBentrySTDinterwordspacing}{\spaceskip=0pt\relax}
\providecommand{\BIBentryALTinterwordstretchfactor}{4}
\providecommand{\BIBentryALTinterwordspacing}{\spaceskip=\fontdimen2\font plus
\BIBentryALTinterwordstretchfactor\fontdimen3\font minus
  \fontdimen4\font\relax}
\providecommand{\BIBforeignlanguage}[2]{{%
\expandafter\ifx\csname l@#1\endcsname\relax
\typeout{** WARNING: IEEEtran.bst: No hyphenation pattern has been}%
\typeout{** loaded for the language `#1'. Using the pattern for}%
\typeout{** the default language instead.}%
\else
\language=\csname l@#1\endcsname
\fi
#2}}
\providecommand{\BIBdecl}{\relax}
\BIBdecl

\bibitem{AMET:AMET219}
A.~W. Wolfe, ``{Social Network Analysis: Methods and Applications},''
  \emph{American Ethnologist}, 1997.

\bibitem{DBLP:conf/sacmat/GeorgiadisMPT01}
C.~K. Georgiadis, I.~Mavridis, G.~Pangalos, and R.~K. Thomas, ``Flexible
  team-based access control using contexts,'' in \emph{SACMAT}, 2001, pp.
  21--27.

\bibitem{Weitzner05creatingthe}
D.~J. Weitzner, J.~Hendler, T.~Berners-lee, and D.~Connolly, \emph{Creating the
  Policy-Aware Web: Discretionary, Rules-based Access for the World Wide
  Web}.\hskip 1em plus 0.5em minus 0.4em\relax IOS Press, 2005.

\bibitem{GWT}
Google, ``{GWT: Google Web Toolkit},'' Available online at
  \url{http://code.google.com/webtoolkit/}.

\bibitem{MySQL}
Oracle, ``Mysql,'' Available online at
  \url{http://www.mysql.com/products/community/}.

\bibitem{DBLP:www/org/w3/http1-1}
R.~T. Fielding, H.~F. Nielsen, and T.~Berners-Lee. (1999) {Internet Draft:
  Hypertext Transfer Protocol - HTTP/1.1}.

\bibitem{Fielding:200:AS}
R.~T. Fielding, ``Architectural styles and the design of network-based software
  architectures,'' Ph.D. dissertation, University of California, Irvine, 2000.

\bibitem{RFC0791}
\BIBentryALTinterwordspacing
J.~Postel, ``{RFC 791}: {Internet Protocol},'' Sep. 1981, obsoletes RFC0760.
  See also STD0005. Status: STANDARD. [Online]. Available:
  \url{ftp://ftp.internic.net/rfc/rfc760.txt,
  ftp://ftp.internic.net/rfc/rfc791.txt,
  ftp://ftp.math.utah.edu/pub/rfc/rfc760.txt,
  ftp://ftp.math.utah.edu/pub/rfc/rfc791.txt}
\BIBentrySTDinterwordspacing

\bibitem{ec2}
{Amazon Web Services, Inc.}, ``{Amazon Elastic Compute Cloud (Amazon EC2),
  Cloud Computing Servers},'' {Available on-line at
  \url{http://aws.amazon.com//ec2/}}.

\bibitem{diaspora}
{DiasporaProject.org}, ``The diaspora project,'' {Available on-line at
  \url{https://joindiaspora.com/}}, 2013.

\bibitem{appleseed}
{The Appleseed Project}, ``{The Appleseed Project - Open Source Social
  Networking},'' {Available on-line at
  \url{http://opensource.appleseedproject.org/}}, 2011.

\bibitem{DBLP:conf/comsnets/ShakimovLCCLLV11}
A.~Shakimov, H.~Lim, R.~C{\'a}ceres, L.~P. Cox, K.~A. Li, D.~Liu, and
  A.~Varshavsky, ``Vis-{\`a}-vis: Privacy-preserving online social networking
  via virtual individual servers,'' in \emph{COMSNETS}, 2011.

\bibitem{supernova}
R.~Sharma and A.~Datta, ``Supernova: Super-peers based architecture for
  decentralized online social networks,'' \emph{COMSNETS}, 2012.

\bibitem{p2p11-social}
G.~Mega, A.~Montresor, and G.~P. Picco, ``Efficient dissemination in
  decentralized social networks,'' in \emph{P2P}.\hskip 1em plus 0.5em minus
  0.4em\relax IEEE, Aug. 2011, CONFERENCE, pp. 338--347.

\bibitem{freenet}
{The FreeNet Project}, ``{Freenet, the free network},'' {Available on-line at
  \url{https://freenetproject.org/}}, 2013.

\bibitem{DBLP:reference/social/DattaBVSR10}
A.~Datta, S.~Buchegger, L.-H. Vu, T.~Strufe, and K.~Rzadca, ``{Decentralized
  Online Social Networks},'' in \emph{Handbook of Social Network Technologies},
  2010.

\bibitem{tbl}
C.~man Au~Yeung, I.~Liccardi, K.~Lu, O.~Seneviratne, and T.~Berners-Lee,
  ``{Decentralization: The Future of Online Social Networking},'' in
  \emph{Proceedings of W3C Workshop on the Future of Social Networking}, 2009.

\bibitem{federatedWiki}
W3C, ``Federated social web incubator group,'' Available online at
  \url{http://www.w3.org/2005/Incubator/federatedsocialweb/wiki/Main_Page}.

\bibitem{oracle}
{Oracle Inc.}, ``{Oracle Social Network},'' {Available on-line at
  \url{http://cloud.oracle.com/mycloud/f?p=service:social:0}}.

\bibitem{yammer}
Yammer, ``{Yammer - The enterprise social network},'' {Available at
  \url{https://www.yammer.com/}}, 2013.

\bibitem{salesforce}
{SalesForce.com Inc.}, ``{SalesForce},'' {Available on-line at
  \url{http://www.salesforce.com/}}.

\bibitem{springerlink}
B.~Carminati and E.~Ferrari, ``{Privacy-Aware Access Control in Social
  Networks: Issues and Solutions},'' in \emph{Privacy and Anonymity in
  Information Management Systems}.\hskip 1em plus 0.5em minus 0.4em\relax
  Springer London, 2010.

\bibitem{BFbook}
F.~Bonchi and E.~Ferrari, Eds., \emph{{Privacy-aware Knowledge Discovery: Novel
  Applications and New Techniques}}.\hskip 1em plus 0.5em minus 0.4em\relax
  {Chapman and Hall/CRC Press}, 2010.

\bibitem{Fong:2011:RAC:1943513.1943539}
P.~W. Fong, ``Relationship-based access control: protection model and policy
  language,'' in \emph{{CODASPY}}, 2011.

\bibitem{DBLP:journals/tissec/CarminatiFP09}
B.~Carminati, E.~Ferrari, and A.~Perego, ``Enforcing access control in
  web-based social networks,'' \emph{ACM Trans. Inf. Syst. Secur.}, 2009.

\bibitem{DBLP:journals/ac/Sandhu98}
R.~S. Sandhu, ``Role-based access control,'' \emph{Advances in Computers},
  vol.~46, pp. 237--286, 1998.

\bibitem{DBLP:conf/rbac/Thomas97}
R.~K. Thomas, ``{Team-based access control (TMAC): a primitive for applying
  role-based access controls in collaborative environments},'' in \emph{ACM
  Workshop on Role-Based Access Control}, 1997, pp. 13--19.

\end{thebibliography}

\end{document}